\documentclass[preprint,showpacs,showkeys,preprintnumbers,amsmath,amssymb,prb]{revtex4}
\usepackage{graphicx, epsfig}
\usepackage{amssymb}
\usepackage{bbm}
\usepackage{subfigure}
\usepackage{sidecap}
\usepackage{mathtools}
\usepackage{relsize}

\begin{document}

\title{Diamagnetic persistent currents for electrons in ballistic billiards subject to a point flux}

\author{\bf Oleksandr Zelyak$^1$}
  \email{zelyak@pa.uky.edu}
\author{\bf Ganpathy Murthy$^{1,2}$}
  \email{murthy@pa.uky.edu}
\affiliation{
$^1$Department of Physics and Astronomy,
University of Kentucky, Lexington, Kentucky 40506, USA\\
$^2$Department of Physics, Harvard University, Cambridge, Massachusetts, 02138
}

\date{\today}

\begin{abstract}
We study the persistent current of noninteracting electrons subject to
a pointlike magnetic flux in the simply connected chaotic Robnik-Berry
quantum billiard, and also in an annular analog thereof. For the
simply connected billiard we find a large diamagnetic contribution to
the persistent current at small flux, which is independent of the flux
and is proportional to the number of electrons (or equivalently the
density since we keep the area fixed). The size of this diamagnetic
contribution is much larger than mesoscopic fluctuations in the
persistent current in the simply connected billiard, and can
ultimately be traced to the response of the angular momentum $l=0$
levels (neglected in semiclassical expansions) on the unit disk to a
pointlike flux at its center. The same behavior is observed for the
annular billiard when the inner radius is much smaller than the outer
one, while the usual fluctuating persistent current and Anderson-like
localization due to boundary scattering are seen when the annulus
tends to a one-dimensional ring. We explore the conditions for the
observability of this phenomenon.
\end{abstract}



\maketitle

%
%
%%%%%%%%%%%%%%%%%%%%%%%%%%%%%%%%%%%%%%%%%%%%%%%%%%%%%%%
%
      \section{Introduction}
%
%%%%%%%%%%%%%%%%%%%%%%%%%%%%%%%%%%%%%%%%%%%%%%%%%%%%%%%
%

A resistanceless flow of electrons can occur in mesoscopic systems if
the linear size $L$ is less than the phase coherence length
$L_{\phi}$. The simplest example of this is a one-dimensional metallic
ring threaded by a magnetic flux $\Phi$. The thermodynamic relation
\begin{equation}
I=-{\partial F\over\partial\Phi}
\end{equation}
defines the persistent current in MKS units. At zero temperature,
which we will focus on, the free energy $F$ can be replaced by the
total ground state energy $E$.

Persistent currents were first predicted to occur in superconducting
rings\cite{paper_Byers_Yang, paper_Bloch, paper_Gunther1969}. It was
later realized that persistent currents exist in normal metallic rings
as well\cite{paper_Buettiker1983, paper_Buttiker_1985}. The phenomenon
is understood most easily at zero temperature for a ring of
noninteracting electrons, where the electronic wavefunction extends
coherently over the whole ring. If the ring is threaded by a
solenoidal flux, all physical properties are periodic in applied
magnetic flux with a period of the flux quantum $\Phi_0=h/e$. A
nonzero flux splits the degeneracy between clockwise and anticlockwise
moving electrons. Upon filling the energy states with electrons, one
finds ground states which have net orbital angular momentum, and net
persistent current. Much experimental work has been carried out on
ensembles of rings/quantum dots\cite{Levy_1993,Marcus_1992} in a flux
as well as on single
metallic\cite{Altshuler_1982,Webb_1985,Chandrasekhar_1985,Chandrasekhar_1991,Deblock2_2002}
or semiconductor quantum
dots/rings\cite{Timp_1987,Wees_1989,Yacoby_1995,Deblock_2002}. The
subject has a long theoretical history as
well\cite{Cheung_1988,Cheung_1989,Ambegaokar_1990,Ambegaokar2_1990,Eckern_1991,Ambegaokar_1991,Schmid_1991,Oppen_1991,Altshuler_1991,Apel_2000}.

In this paper we investigate the persistent current of noninteracting
electrons in quantum billiards subject to a point flux. Related
semiclassical calculations have been carried out in the past for
regular (integrable in the absence of
flux)\cite{Oppen_1993,Reimann_1996,Jalabert_1996,Richter_1996,Narevich_2000,Howls_2001}
and chaotic billiards\cite{Oppen_1993,Jalabert_1996,Richter_1996,Kawabata_1999}. Numerics have previously
been performed on these systems as
well\cite{Ree_1999,Narevich_2001}. We carry out calculations on the
simply connected chaotic Robnik-Berry
billiard\cite{Berry_1977,Robnik_1984,paper_Berry_Robnik,paper_Bruus_Stone_arxiv,paper_Bruus_Stone_prb}
obtained by deforming the boundary of the integrable disk, and on an
annular analog which we call the Robnik-Berry annulus. The ratio of
the inner $r$ to the outer radius $R$ of the annulus ($\xi=r/R$) plays
an important role in our analysis, and allows us to go continuously
between the simply connected chaotic two-dimensional billiard and a
(effectively disordered) quasi-one-dimensional ring.

Our main result is that there is a large {\it diamagnetic and
flux-independent} contribution to the persistent current for
$|\Phi|\ll\Phi_0$ in the simply connected billiard which is
proportional to the number of particles, and overwhelms the mesoscopic
fluctuations which have been the focus of previous
work\cite{Cheung_1988,Cheung_1989,Schmid_1991,Oppen_1991,Altshuler_1991,Oppen_1993,Reimann_1996,Jalabert_1996,Richter_1996,Narevich_2000,Howls_2001,Kawabata_1999,Ree_1999,Narevich_2001}. This
arises from angular momentum $l=0$ states in the integrable disk,
which respond diamagnetically, with energy increasing {\it linearly}
with the point flux, for small flux. This behavior is robust under the
deformation of the boundary which makes the dynamics chaotic. As $\xi$
increases from zero, this contribution to the persistent current
persists for typical $\Phi/\Phi_0\simeq1$, but smoothly decreases in
magnitude and becomes negligible for $\xi\to1$. The precise $\xi$ at
which the diamagnetic contribution to the persistent current becomes
equal to the typical fluctuating paramagnetic contribution depends on
the electron density. For $\xi\ne0$ and very tiny flux the diamagnetic
contribution to the persistent current varies linearly with
$\Phi/\Phi_0$ (see below).

This diamagnetic contribution seems to have been missed in previous
work, to the best of our knowledge. The reason is that the
semiclassical approximation becomes asymptotically exact as the energy
tends to infinity, and in this limit, the spectral density of $l=0$
states vanishes. Thus, $l=0$ states are explicitly
disregarded\cite{Oppen_1993,Jalabert_1996,Richter_1996,Kawabata_1999}
in the semiclassical approach, since they do not enclose flux. It has
been noted in the past that diffraction effects necessitate an
inclusion of $l=0$ states in the sum over periodic orbits on the
integrable disk\cite{Reimann_1996}, but the connection to persistent
currents was not made.

It should be emphasized that since the total persistent current is a
sum over the contributions of all levels, the diamagnetic contribution
we uncover exists even at very large energies, where the levels at the
Fermi energy are well approximated by semiclassics.

The robustness of the diamagnetic contribution to the persistent
current under deformation can be understood as follows: In the chaotic
billiard, each state at a particular energy is roughly a linear
combination of states of the disk within a Thouless energy ($E_T\simeq
\hbar v_F/L$, where $L$ is the linear size of the billiard) of its
energy. When the Fermi energy $E_F$ greatly exceeds $E_T$, the
contribution of the occupied states does not change much when the
boundary is deformed and chaos is introduced.

This behavior appears similar to, but is different from Landau
diamagnetism\cite{Nakamura_1988} in a finite system, which is a
response to a uniform magnetic field. The primary difference is that
the orbital magnetization (proportional to the persistent current) in
Landau diamagnetism is proportional to the field itself (because the
energy goes quadratically with the field strength), whereas the effect
we describe is independent of the flux for small flux in the simply
connected Robnik-Berry billiard (because the energy goes linearly with
the flux).  In the Robnik-Berry annulus with $\xi\to0$, the energy
rises quadratically with the flux for very tiny flux $\Phi\ll
\Phi_0/\log{N\xi^{-2}}$, but crosses over to the linear behavior
characteristic of the simply connected system for larger $\Phi$. Since
the flux is pointlike, and in the annular case nonzero only where the
electron wavefunctions vanish, the entire effect is due to
Aharanov-Bohm quantum interference. Since the effect is primarily
caused by levels deep below $E_F$, experimental detection is feasible
only through the total magnetization, and not by conductance
fluctuations which are sensitive to the levels within the Thouless
shell (lying within $E_T$ of $E_F$). Previous samples have been
subjected to a uniform field rather than a point
flux\cite{Altshuler_1982,Webb_1985,Chandrasekhar_1985,Chandrasekhar_1991,Timp_1987,Wees_1989,Yacoby_1995},
and anyway the ring samples have $\xi$ too large for this effect to be
seen. However, we believe that experiments can be designed to observe
this effect.

The plan of the paper is as follows: In Section II we describe the
method we use to calculate the spectrum, and present analytical
expressions and numerical results for persistent current in the disk
and simply connected Robnik-Berry billiards. In Section III we
generalize the method to  the annulus and present our
results. Conclusions and implications are presented in Section IV.

%
%%%%%%%%%%%%%%%%%%%%%%%%%%%%%%%%%%%%%%%%%%%%%%%%%%%%%%%
%
      \section{The simply connected Robnik-Berry billiard}
%
%%%%%%%%%%%%%%%%%%%%%%%%%%%%%%%%%%%%%%%%%%%%%%%%%%%%%%%
%

We begin by briefly describing the procedure to obtain the energy levels $\epsilon_k$  
within the  billiard, which leads to the persistent current:
\begin{equation}
  I=\sum_{k} I_k, \ \ \ \  I_k = -\frac{\partial \epsilon_k}{\partial \Phi},
\label{eq:04}
\end{equation}

We work with the Robnik-Berry billiard\cite{Robnik_1984,paper_Berry_Robnik}, which
is obtained from the unit disk by conformal transformation. The
original problem of finding energy levels of electron in the domain
with complicated boundaries is reduced to a problem where the electron
moves in the unit disk in a fictitious potential introduced by
the following conformal transformation; 
\begin{equation} 
w(z) = \frac{z + bz^2 + ce^{i\delta} z^3}{\sqrt{1 + 2b^2 + 3c^2}}
\end{equation} 
where $w=u+iv$ represents the coordinates in the laboratory coordinate
system, and $z=x+iy$ are the conformally transformed coordinates
(details are in Appendix A).  The parameters $b$, $c$, and $\delta$
control the shape of the original billiard, and for the values we use,
the classical dynamics is mixed, but largely chaotic. It is also
straightforward to introduce a point flux which penetrates the center
of the unit
disk\cite{Robnik_1984,paper_Berry_Robnik,paper_Bruus_Stone_arxiv,paper_Bruus_Stone_prb}
(after the conformal transformation).

 We find 600 energy levels for regular and chaotic billiards for
different values of parameter $\alpha$ that controls magnetic flux
coming through billiards. Only the lowest 200 levels are actually used
in further calculations, since the higher levels become increasingly
inaccurate\cite{paper_Bruus_Stone_arxiv,paper_Bruus_Stone_prb}. The
persistent current is obtained as a numerical derivative of the ground
state energy for a given number of electrons.

\begin{figure*}
  \centerline{ 
    \mbox{\includegraphics[width=3.5in]{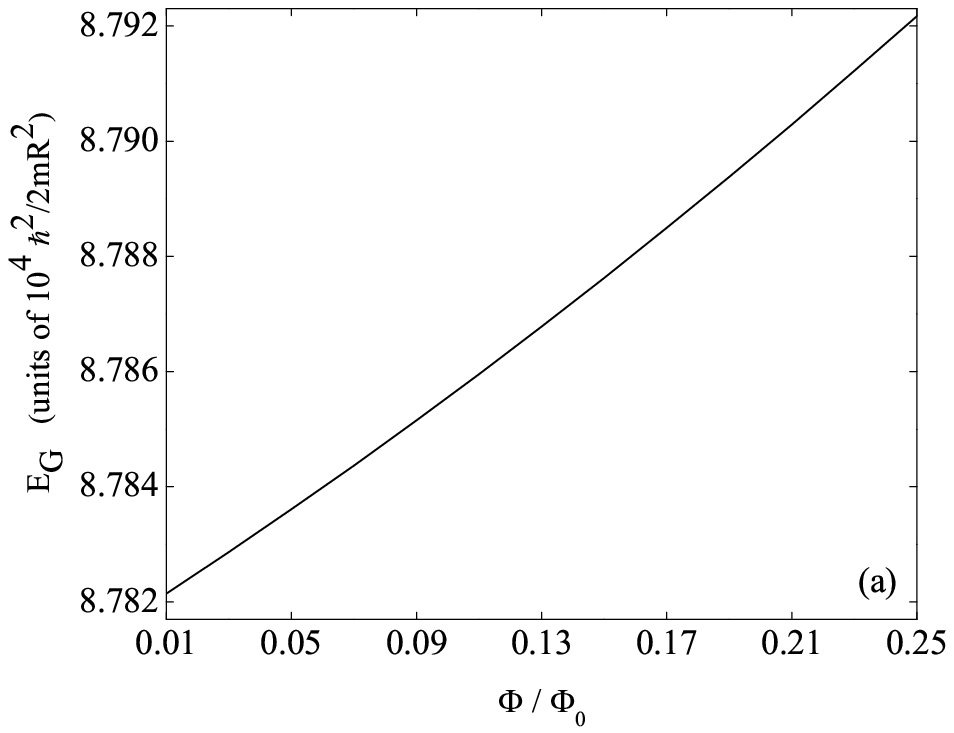}}
    \mbox{\includegraphics[width=3.33in]{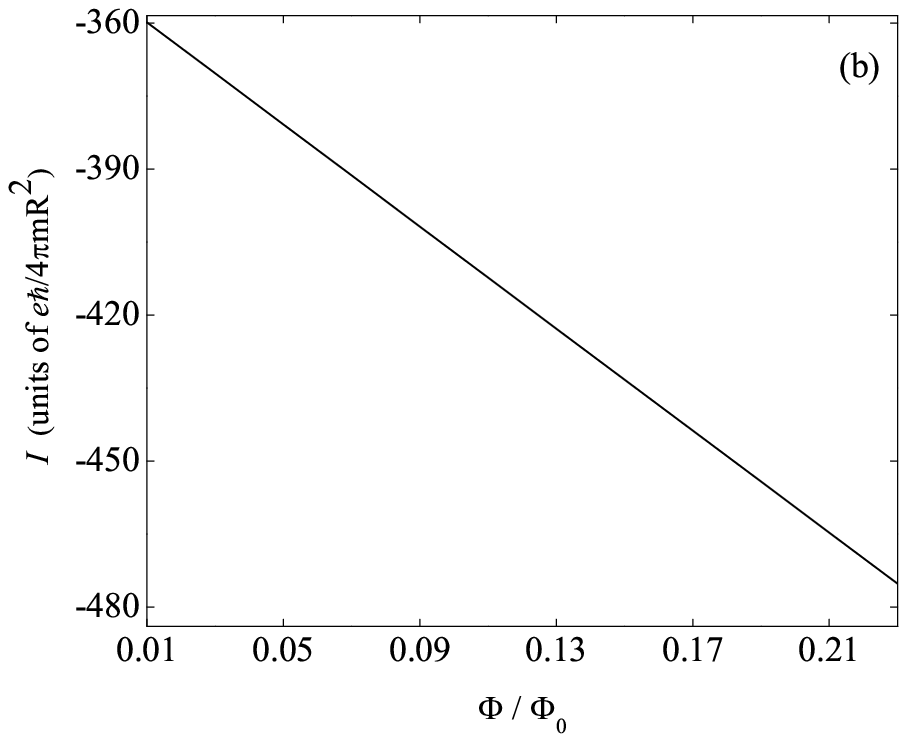}}
  }
  \centerline{
    \mbox{\includegraphics[width=3.5in]{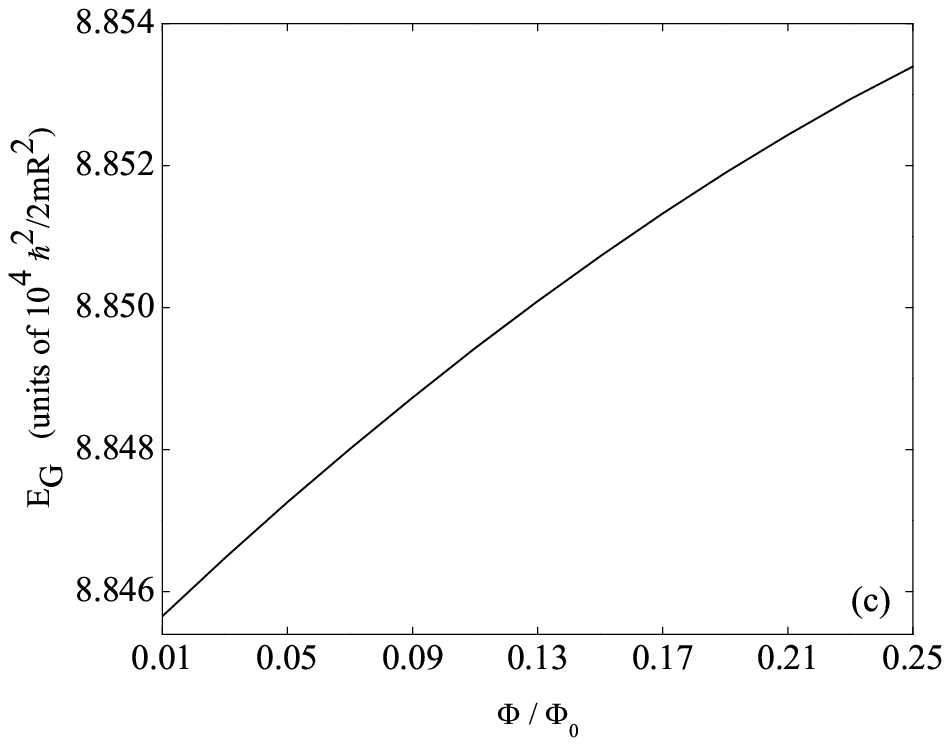}}
    \mbox{\includegraphics[width=3.33in]{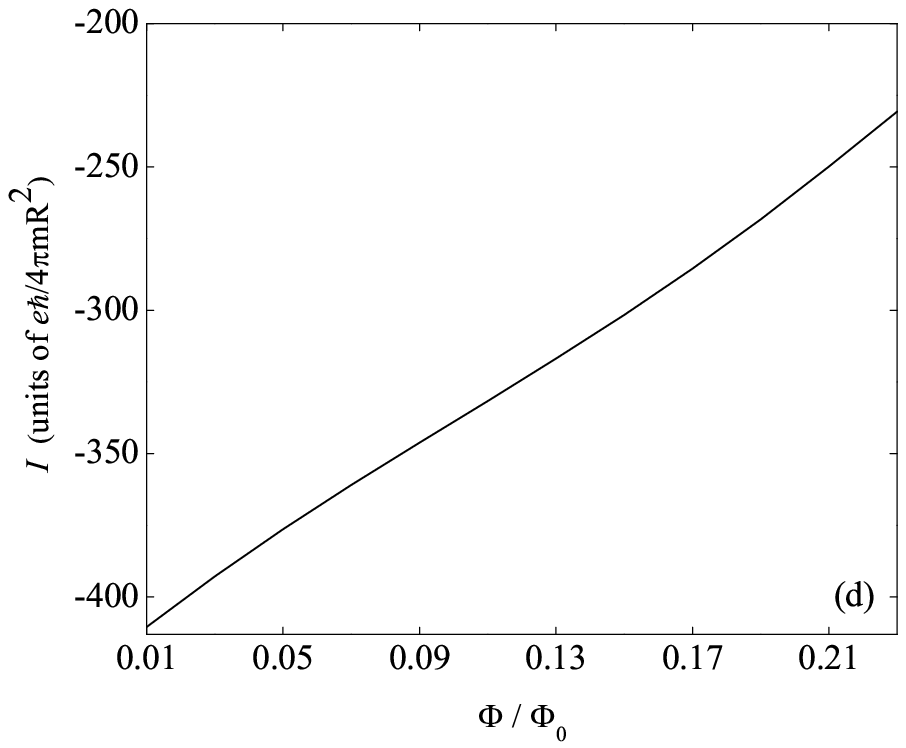}}
  }
  \caption{Ground state energy $E_G$ in units of
           $10^4\times\frac{\hbar^2}{2mR^2}$, and persistent current
           $I$ in units of $\frac{e\hbar}{4\pi mR^2}$ as a function of
           dimensionless flux for the regular disk (panels a,b) and
           the simply connected chaotic billiard (panels c,d). The
           results are for 200 particles. }
  \label{Eg_and_I_vs_alpha_DISK}
  \end{figure*}

 For the unit disk billiard the ground state energy
$E_G$ has a non zero slope as $\alpha \rightarrow 0$. Thus there is a
persistent current in the system for arbitrarily small magnetic flux
(see Fig. \ref{Eg_and_I_vs_alpha_DISK}).

%----------------------------------------------------------------------------
%------------------ Figures: Eg and I versus alpha for disk -----------------
%----------------------------------------------------------------------------

%-----------------------------------------------------------------------------

  Qualitatively this  behavior can be understood as follows. In
the absence of magnetic field energy levels corresponding to orbital
quantum numbers $\pm l$, are degenerate.  A nonzero $\Phi$ lifts the
degeneracy and for small $\alpha$ the two $\pm l$ levels have slopes
that are equal in magnitude and opposite in sign. Thus, as long as
both are occupied, these levels do not contribute to the net
persistent current $I$.  The only nonzero contribution comes from
levels with $l=0$.

  For the unit disk the expression for the persistent current can be
derived analytically for small values of magnetic flux. At zero
temperature the persistent current due to $k$th level is $I_k = - \partial
\epsilon_k / \partial \Phi$ ($\epsilon_k$ is a dimensionless energy,
and $I_k$ is persistent current divided by the energy unit
$\hbar^2/2mR^2$; see appendix \ref{apnx:A} for notations). For the
unit disk, the energy levels are found from the quantization
condition:
\begin{equation}
 J_{|l-\alpha|}(\gamma_{|l-\alpha|,n}) = 0, \qquad \epsilon_k = \gamma^2_n(|l-\alpha|).
\label{eq:05}
\end{equation}
  Then from Eq. \eqref{eq:04}, persistent current caused by $k$th level is:
\begin{equation}
  I_k = - \frac{2e}{h} \gamma_n(|l_k-\alpha|)
         \frac{\partial \gamma_n(|l_k-\alpha|)}{\partial \alpha}.
\label{eq:10}
\end{equation}

 To find $\partial\gamma/\partial\alpha$ we differentiate Eq. \eqref{eq:05}:
\begin{equation}
  \frac{\partial J_{\nu}(\gamma)}{\partial \alpha}
    = 
      \frac{\partial J_{\nu}(\gamma)}{\partial \nu}
      \frac{\partial \nu}{\partial \alpha}
    +
      \frac{\partial J_{\nu}\left(\gamma \right)}{\partial \gamma}
      \frac{\partial \gamma}{\partial \alpha} = 0.
\label{eq:06}
\end{equation}

For $l=0$ levels, $\nu = |l-\alpha| = \alpha$. In $\alpha \rightarrow 0$ limit, 
for the derivatives of Bessel function one gets:
\begin{equation}
 \begin{split}
     \frac{\partial J_{\nu}(\gamma)}{\partial \nu}\biggl|_{\nu=0}
  &=  \frac{\pi}{2} N_0(\gamma),\\
     \frac{\partial J_{\nu}(\gamma)}{\partial \gamma}\biggl|_{\nu=0}
  &=  -J_1(\gamma).
 \end{split}
\label{eq:07}
\end{equation}

 Combining Eqs. \eqref{eq:06} and \eqref{eq:07} and using relation
that for $\gamma \gg 1$ Bessel function $N_0(\gamma) \approx
J_1(\gamma)$ (this approximation works well already for the first root
of Eq. \eqref{eq:05}), we find
$\frac{\partial\gamma}{\partial\alpha}|_{\alpha=0}=\frac{\pi}{2}$,
which leads to:
\begin{equation}
    I = -\frac{\pi e}{h}\sum_n \gamma_n(0), 
\label{eq:disk_pers_cur}
\end{equation} 
\noindent  where summation is over the levels with orbital quantum number $l=0$.

For large argument values (which is the same as large energies) the
quantization condition \eqref{eq:05} for the unit disk becomes
$\cos(\gamma_n - \pi\alpha/2 - \pi/4) = 0$, with roots:
\begin{equation}
  \gamma_n = \pi\alpha/2 + \pi/4 + \pi(2n+1)/2.
  \label{eq:disk_roots}
\end{equation}

With the energy being measured in $\hbar^2/2mR^2$ units, the Fermi
wave vector is $k_F = \gamma_{max} \approx \pi n_{max}$, where
$n_{max}$ denotes the largest $l=0$ level. With disk area equal to
$\pi$ ($R=1$), the number of particles in the system is $N = (\pi
n_{max}/2)^2$. This allows us to find the dependence of the persistent
current on number of particles in the system in $\alpha \rightarrow 0$
limit.
\begin{equation}
 I = -\frac{e\pi}{h} \sum_n (\frac{3\pi}{4} + \pi n) \approx -\frac{e\pi^2}{2h} n^2_{max}
   = -\frac{e}{h}2N,
  \label{eq:I_vs_N_disk}
\end{equation}

\noindent where we neglected a subleading term proportional to
$n_{max}$. We remind the reader that the physical persistent current
is the expression in formula \eqref{eq:I_vs_N_disk} divided by energy
unit $\hbar^2/2mR^2$.

\begin{figure*}
  \centerline{ 
    \mbox{\includegraphics[width=3.5in]{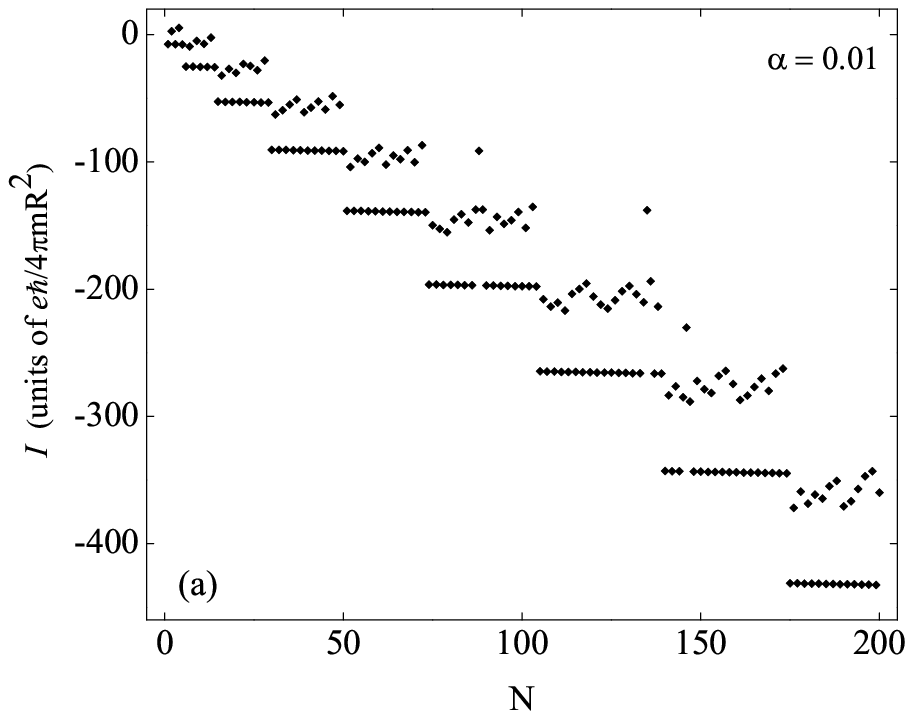}}
    \mbox{\includegraphics[width=3.45in]{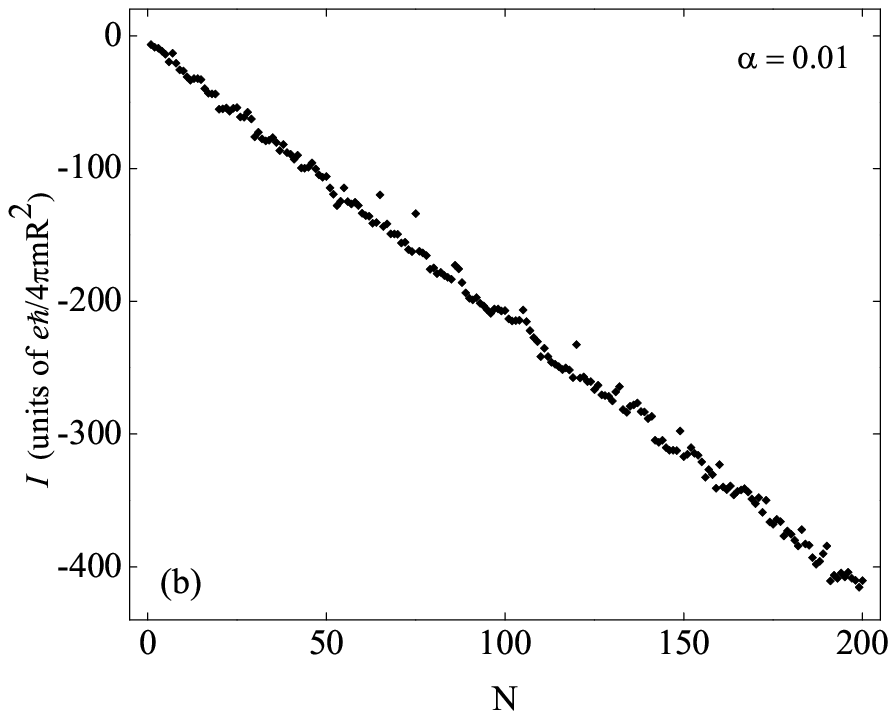}}
  }
  \caption{Persistent current $I$ vs. number of particles $N$ for regular disk (a)
           and chaotic disk (b) for the value of reduced magnetic flux $\alpha = 0.01$.}
  \label{I_vs_N_alpha_.01_DISK}
  \end{figure*}

In Fig. \ref{I_vs_N_alpha_.01_DISK} the persistent current $I$ is
plotted against the number of particles $N$ for magnetic flux $\alpha
= 0.01$. The behavior of the current is consistent with
Eq. \eqref{eq:I_vs_N_disk}. That is, for small magnetic flux it is
proportional to $2N$. For the regular disk
(Fig. \ref{I_vs_N_alpha_.01_DISK}a) the persistent current is a set of
consecutive steps. Each step appears when the next $l=0$ level is
added to the system.  The length of the steps is equal to the number
of $l\ne 0$ levels between two adjacent levels with zero orbital
quantum number. As one particle is added to the $l\ne 0$ level, it
results in persistent current jump. The next level has opposite slope,
and once it is occupied, cancels the contribution of the previous
$l\ne 0$ level to the net persistent current. This explains the noise
above each step in Fig. \ref{I_vs_N_alpha_.01_DISK}a. In addition,
each step has a small inclination which is due to the fact that the
$l\ne 0$ levels do not cancel each other exactly when $\Phi\ne 0$. For
larger magnetic flux the steps become more inclined.

%----------------------------------------------------------------------------
%-------------- Figures: I versus N for regular and chaotic disk ------------
%----------------------------------------------------------------------------
  
%-----------------------------------------------------------------------------

To see that levels with $l\ne 0$ do not contribute to persistent current
at weak magnetic flux we simply note that derivative of $\gamma$:
\begin{equation}
      \frac{\partial \gamma_n}{\partial \alpha}\biggl|_{\alpha=0}
   =  - \frac{\partial J_{\nu}(\gamma_n)}{\partial \nu} \frac{\partial \nu}{\partial \alpha}
        \biggm / \frac{\partial J_{\nu}(\gamma_n)}{\partial \gamma_n} \biggl|_{\alpha = 0}
\label{eq:09}
\end{equation}
   is an odd function of $l$. In the $\alpha \rightarrow 0$ limit, the
root $\gamma_n(\nu)$ and the derivatives of $J_{\nu}(\gamma_n)$ in
Eq. \eqref{eq:09} are even functions of $l$, and $\partial \nu /
\partial \alpha$ is odd. As a result, the whole expression is odd
function of $l$, which proves the cancellation of $\pm l$ levels in
Eq. \eqref{eq:10}.

  For the chaotic simply connected Robnik-Berry billiard each
eigenstate is a superposition of all $l$-states of regular disk (see
Eq. \eqref{eq:03}), mostly within a Thouless shell of its
energy. Assuming that states with $\pm l$ enter this superposition
with equal probability over the ensemble due to the chaotic nature of
motion, one can conclude that the ensemble-averaged contribution of
these levels to the net current is zero. However, as seen in
Fig. \ref{I_vs_N_alpha_.01_DISK}b, the mesoscopic fluctuations due to
the $l\ne0$ levels are overwhelmed by the diamagnetic contribution
linear in $N$ for small $\alpha$.

%
%%%%%%%%%%%%%%%%%%%%%%%%%%%%%%%%%%%%%%%%%%%%%%%%%%%%%%%
%
      \section{The Robnik-Berry annulus}
% 
%%%%%%%%%%%%%%%%%%%%%%%%%%%%%%%%%%%%%%%%%%%%%%%%%%%%%%%
%

  Now we turn our attention to annular billiard. Here there is an
additional parameter $\xi$, which is the ratio of the inner radius $r$ to the
outer radius $R$ of the regular annulus in the (conformally transformed)
$z$ plane.  By varying $\xi$ we are able to smoothly go from the
simply connected Robnik-Berry billiard to an (effectively) disordered
ring in the limit $\xi\to1^-$.

First consider the disk limit $\xi\to0$. We can derive an analytical
expression for $I$ when $\xi$ is small enough that
$\xi\gamma_n \ll 1$, and $\gamma_n \gg 1$.
For a regular annulus with $R=1$ energy quantization follows from the Dirichlet
boundary condition:
\begin{equation}
  J_{\nu}(\gamma_{n}) N_{\nu}(\gamma_{n}\xi) - 
                J_{\nu}(\gamma_{n}\xi) N_{\nu}(\gamma_{n}) = 0.
  \label{eq:quant_cond_ann}
\end{equation}

\noindent Here $n$ numerates root at fixed angular momentum
$\nu$. We use the large and small argument expansions
for Bessel functions to obtain for the  $l=0$ levels:
\begin{equation}
  \cot(\gamma_n - \frac{\pi\alpha}{2} - \frac{\pi}{2}) = 
   \frac{\dfrac{1}{\Gamma(1+\alpha)}\bigg(\dfrac{\gamma_n\xi}{2}\bigg)^{\alpha} }{
          \dfrac{\cot(\alpha\pi)}{\Gamma(1+\alpha)} 
                \bigg(\dfrac{\gamma_n\xi}{2}\bigg)^{\alpha} -
          \dfrac{\Gamma(\alpha)}{\pi} \bigg(\dfrac{\gamma_n\xi}{2}\bigg)^{-\alpha}   }.
\end{equation}

  We express the  roots for the annulus as a small deviation from the roots for the disk, which we denote $\gamma_n^{(d)}$; 
$\gamma_{n} = \gamma^{(d)}_n + \delta\gamma_n$ with 
$\gamma^{(d)}_n = \alpha\pi/2 + \pi/4 + \pi(2n+1)/2$. Approximating $\cot(\alpha\pi)$
by $1/(\alpha\pi)$, for small values of $\delta\gamma_n$ we find:
\begin{equation}
  \delta\gamma_n = - \frac{\alpha\pi}{2}\bigg[1+\coth\bigg(\alpha\ln\frac{\gamma_n\xi}{2}\bigg)\bigg].
  \label{eq:var_gamma_annul}
\end{equation}

   In Eq. \eqref{eq:var_gamma_annul}, for small $\alpha$, $\gamma_n$
under logarithm can be safely replaced by its value for the disk
$\gamma^{(d)}_n$. Then roots for the annulus are:
\begin{equation}
  \gamma_n = \frac{\pi}{4} + \frac{\pi}{2}(2n + 1) 
              - \frac{\pi}{2}\alpha\coth\bigg(\alpha\ln\frac{\gamma^{(d)}_n\xi}{2}\bigg).
  \label{eq:annul_root}
\end{equation}

  One can now take various limits of Eq. \eqref{eq:annul_root}.  To
recover Eq.  \eqref{eq:disk_roots} for the disk roots, we keep
magnetic flux $\alpha$ fixed and take the limit $\xi \rightarrow
0$. As one can see from Eq. \eqref{eq:annul_root}, convergence to the
disk limit is slow due to the logarithm, and occurs only for $\alpha\gg 1/\log{(\gamma^{(d)}_n\xi/2)}$.

  Another limit of interest is to keep $\xi$ fixed and obtain behavior
of roots $\gamma_n$ for small $\alpha$. For small $\alpha\ll
1/\log{(\gamma^{(d)}_n\xi/2)}$, the roots $\gamma_n$ with $l=0$ vanish
quadratically with $\alpha$. Expanding the $\coth$ function in
Eq. \eqref{eq:annul_root}, we obtain:
\begin{equation}
  \gamma_n = \frac{\pi}{4} + \frac{\pi}{2}(2n+1) - \frac{\pi}{2}
   \bigg( 1 + \frac{\alpha^2}{3}\ln^2 \dfrac{\gamma^{(d)}_n\xi}{2} \bigg)
   \ln^{-1}\dfrac{\gamma^{(d)}_n\xi}{2}.
  \label{eq:annul_rut_smal_alpha}
\end{equation}

which leads to the persistent current: 
\begin{equation}
  I \approx \frac{2\pi e\alpha}{3h}\sum_{n}
  \bigg[\bigg(\frac{\pi}{4} + \frac{\pi}{2}(2n+1)\bigg)\ln\dfrac{\gamma^{(d)}_n\xi}{2} - 
         \frac{\pi}{2} \bigg].
\end{equation}

  Rough estimation of this sum with help of the Euler-MacLaurin formula gives:
\begin{equation}
  I \approx \frac{\pi^2e\alpha}{3h}n^2_{max} \ln\frac{n_{max}\xi\pi}{2\sqrt{\textrm{e}}},
\end{equation}

\noindent where we kept only terms proportional to $n^2_{max}$ and
$\textrm{e}=2.71828...$ inside the logarithm denotes Euler's number and not the electronic
charge.

   Using the relation $N = (\pi n_{max}/2)^2$ (for small values of
$\xi$ the density of states for the annulus and the disk are
practically the same), the persistent current becomes for small
$\alpha\ll 1/\log{(\gamma^{(d)}_n\xi/2)}$:
\begin{equation}
  I = I^{(d)} \frac{\alpha}{3}\bigg|\ln\frac{N\xi^2}{\textrm{e}}\bigg|, \hspace{0.5cm} 
      I^{(d)} = -\frac{e}{h}\,2N.
  \label{eq:current_in_annulus}
\end{equation}

   To approach the limit of a one-dimensional ring, where $\gamma_n
\gg 1$ and $\gamma_n\xi \gg 1$, we return to quantization condition
\eqref{eq:quant_cond_ann} and use the following large argument
expansion for Bessel functions:
\begin{align}
  \begin{split}
    J_{\nu}(z) &\approx \sqrt{\frac{2}{\pi z}}\bigg( \cos(z - \frac{\pi\nu}{2} - \frac{\pi}{4}) 
       - \sin(z - \frac{\pi\nu}{2} - \frac{\pi}{4}) \,\frac{\nu^2 - 1/4}{2z} \bigg), \\
    N_{\nu}(z) &\approx \sqrt{\frac{2}{\pi z}}\bigg( \sin(z - \frac{\pi\nu}{2} - \frac{\pi}{4}) 
       + \cos(z - \frac{\pi\nu}{2} - \frac{\pi}{4}) \,\frac{\nu^2 - 1/4}{2z} \bigg).
  \end{split}
 \label{eq:ring_bessle_expans}
\end{align}

  We use formulas \eqref{eq:ring_bessle_expans} and quantization condition 
\eqref{eq:quant_cond_ann} to get a new equation for roots:
\begin{equation}
  \sin(\gamma_n\sigma) - \cos(\gamma_n\sigma)\,\frac{\nu^2 - 1/4}{2\gamma_n\xi}\,\sigma = 0,
  \label{eq:new_quant_ring_condit}
\end{equation}

\noindent where we ignore the term proportional $1/\gamma^2_n$. The
quantity $\sigma = 1 - \xi$, is assumed to be much less than
unity. For sufficiently small $\sigma$ one can drop the second term in
Eq. \eqref{eq:new_quant_ring_condit} and get $\gamma_n\sigma = \pi
n$. To find corrections to this expression we assume that
$\gamma_n\sigma = \pi n + \eta$ with $\eta= \frac{\nu^2 - 1/4}{2\pi
n}\,\sigma^2\ll 1$ and plug it in Eq. \eqref{eq:new_quant_ring_condit}
to obtain the solutions of quantization condition
\eqref{eq:new_quant_ring_condit} are:
\begin{equation}
  \gamma_n = \frac{\pi n}{\sigma} + \frac{\nu^2 - 1/4}{2\pi n}\,\sigma, \hspace{0.5 cm}
           \nu = |l - \alpha|. 
\end{equation}

  The energy spectrum for the annulus in this limit is:
\begin{equation}
  \epsilon_{n,l} = \gamma^2_n \approx \bigg( \frac{\pi n}{\sigma} \bigg)^2 +
       (\nu^2 - 1/4).
  \label{eq:ring_energy}
\end{equation}

  The first term in Eq. \eqref{eq:ring_energy} denotes the radial
kinetic energy and diverges in $\sigma \rightarrow 0$ limit. This
divergence can be absorbed into the chemical potential for the $n=1$
radial state. The difference between energy levels with radial quantum
numbers $n$ is of the order $n(\pi/\sigma)^2$. For
$\sigma\to0\Rightarrow\xi\to1$ one can assume that all the levels of
interest have the radial quantum number $n=1$, and are labelled only
by orbital quantum number $l$. Since our diamagnetic persistent
current arises from a large number $\propto \sqrt{N}$ of $l=0$ levels,
it is clear that it vanishes in the limit of a ring.

 It is straightforward to show that for a regular annulus the 
contributions of $\pm l$ levels also cancel each other for
small values of $\alpha$. However, levels with $l=0$ have zero
slope when $\alpha \rightarrow 0$.  To show this one takes the  derivative
of quantization condition \eqref{eq:quant_cond_ann}:
\begin{multline}
   \dot{J}_{\nu}(\gamma_n\xi) N_{\nu}(\gamma_n) + J_{\nu}(\gamma_n\xi) \dot{N}_{\nu}(\gamma_n)
 - \dot{J}_{\nu}(\gamma_n) N_{\nu}(\gamma_n\xi) - J_{\nu}(\gamma_n) \dot{N}_{\nu}(\gamma_n\xi) \\
 + \frac{\partial \gamma_n}{\partial \alpha} \biggl[ 
     \xi J^{'}_{\nu}(\gamma_n\xi) N_{\nu}(\gamma_n)
  +  J_{\nu}(\gamma_n\xi) N^{'}_{\nu}(\gamma_n)
  -  J^{'}_{\nu}(\gamma_n) N_{\nu}(\gamma_n\xi)
  -  \xi J_{\nu}(\gamma_n) N^{'}_{\nu}(\gamma_n\xi) \biggr] = 0,
\label{eq:12}
\end{multline}
  where $\dot{A}_{\nu}(z) = \partial A_{\nu}(z)/\partial \nu$, and
$A^{'}_{\nu}(z) = \partial A_{\nu}(z)/\partial z$. When $\alpha \rightarrow 0$,
derivatives of Bessel functions become $\dot{J}_{\nu}(z) = \pi N_0(z)/2$,
$\dot{N}_{\nu}(z) = -\pi J_0(z)/2$, $J^{'}_{\nu}(z) = -J_1(z)$, and
$N^{'}_{\nu}(z) = -N_1(z)$. Then all terms outside square brackets in Eq. \eqref{eq:12}
cancel each other. The expression inside brackets
in general has a non-zero value, which means $\partial \gamma_n/\partial \alpha = 0$. 

%----------------------------------------------------------------------------
%--------- Figures: I versus \alpha for regular and chaotic ANNULUS ---------
%----------------------------------------------------------------------------

  \begin{figure*}
  \centerline{
    \mbox{\includegraphics[width=3.5in]{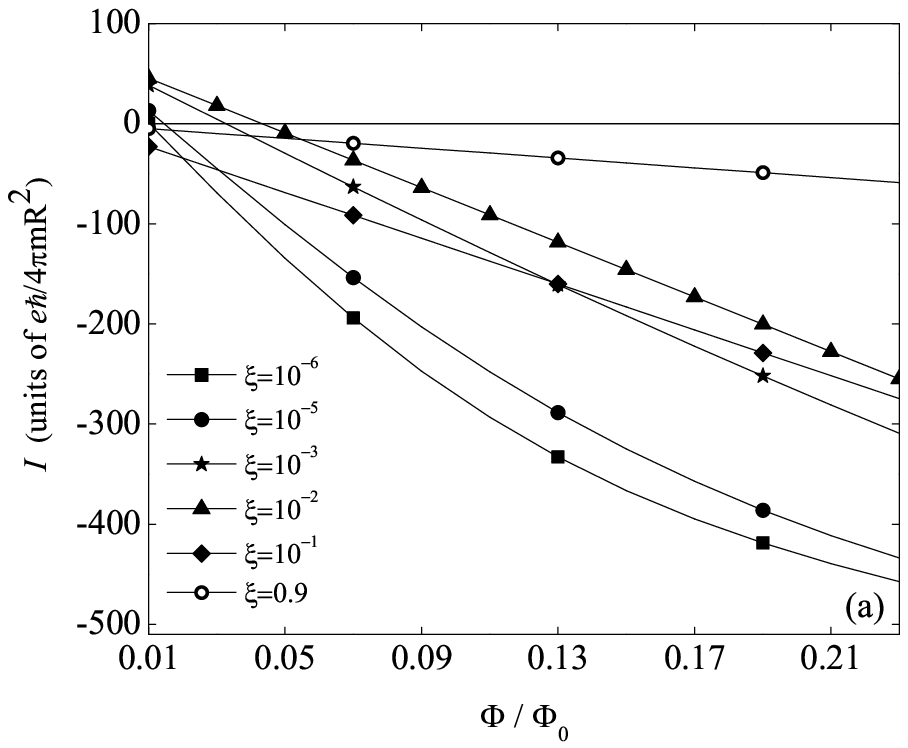}}
    \mbox{\includegraphics[width=3.5in]{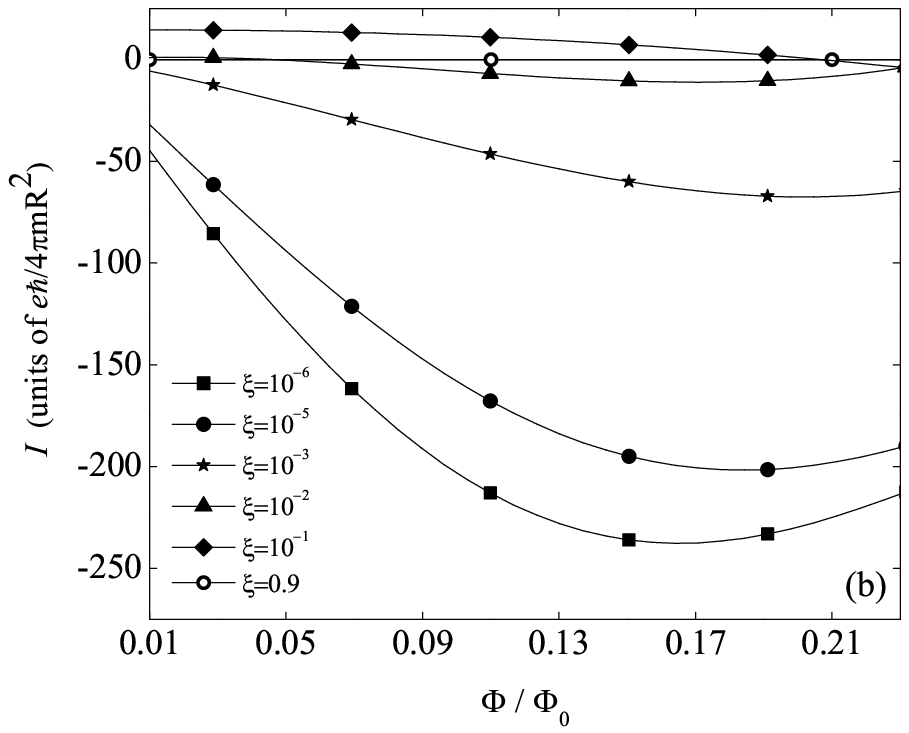}}
  }
  \caption{Persistent current $I$ vs. reduced magnetic flux $\alpha$
            for several values of $\xi$ for $N=200$ particles. Figure
            (a) represents current for regular annulus normalized to
            the same density of states (same area) for different
            values of $\xi$. The current for the chaotic annulus is
            depicted in Figure (b). }
  \label{I_vs_alpha__ANNULUS}
  \end{figure*}

%-----------------------------------------------------------------------------

 In Fig. \ref{I_vs_alpha__ANNULUS} the persistent current in the
annular billiard is depicted for different values of the aspect ratio 
$\xi$. To facilitate the comparison between different values of $\xi$,
we keep the area of the annulus the same, thus keeping the average
density of states the same. For a regular annulus
(Fig. \ref{I_vs_alpha__ANNULUS}a) for small values of flux, the
current is a linear function of $\alpha$. As $\xi$ gets smaller, the
diamagnetic contribution to the persistent current increases. This
behavior is consistent with Eq. \eqref{eq:current_in_annulus} that
shows linear dependence on $\alpha$ and slow growth as $\xi\to0$.

  In the regular annulus, for $\xi$ close to unity the behavior of the
persistent current is close, but not identical, to that of a 1D
ring. Even for $\xi = 0.9$ there  exist several states with
$l=0$, which means that our billiard is not purely a 1D ring. The
effect of these states on the persistent current is not entirely
trivial.  For a fixed number of particles in the system, as $\xi$
changes, the number of $l=0$ levels also changes. As the next $l=0$
level is added (or expelled), the current experiences a jump. The
magnitude of this jump is large enough for small $\alpha$ to make
current to be positive. For larger $\alpha$ the current remains
diamagnetic.

  In the distorted annulus (Fig. \ref{I_vs_alpha__ANNULUS}b) the
persistent current is a linear function of $\alpha$ for small
$\alpha$. For larger magnetic flux one observes nonlinear behavior
that can be attributed to level repulsion in the chaotic billiard.

\begin{figure*}
  \centerline{
    \mbox{\includegraphics[width=3.5in]{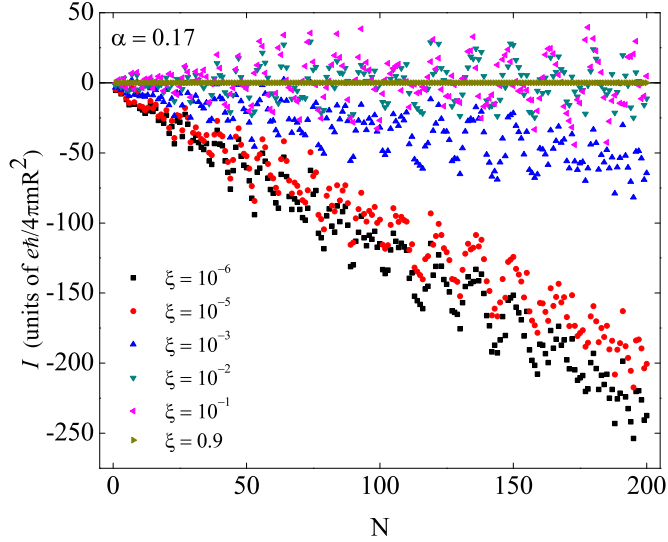}}
  }
  \caption{Persistent current $I$ in distorted annulus vs. number of particles $N$ for
           several values of $\xi$. magnetic flux $\alpha = 0.17$.}
  \label{I_vs_N_xi_all_chaot_ANNULUS}
  \end{figure*}

  The dependence of the persistent current in the annulus on the number of
particles $N$ at fixed $\alpha$ is similar to that in the simply connected 
billiard. At small $\xi$ the persistent current in the regular annulus is a
staircase-like function. For the distorted annulus the numerics  are
scattered around a straight line (see Fig. \ref{I_vs_N_xi_all_chaot_ANNULUS}).

%----------------------------------------------------------------------------
%--------- Figures: I versus N for chaotic ANNULUS at multiple xi; ----------
%----------------------------------------------------------------------------

%-----------------------------------------------------------------------------

%----------------------------------------------------------------------------
%--------- Figures: I versus N for chaotic ANNULUS at xi = 0.9; Localization --------
%----------------------------------------------------------------------------

  \begin{figure*}
  \centerline{
    \mbox{\includegraphics[width=3.5in]{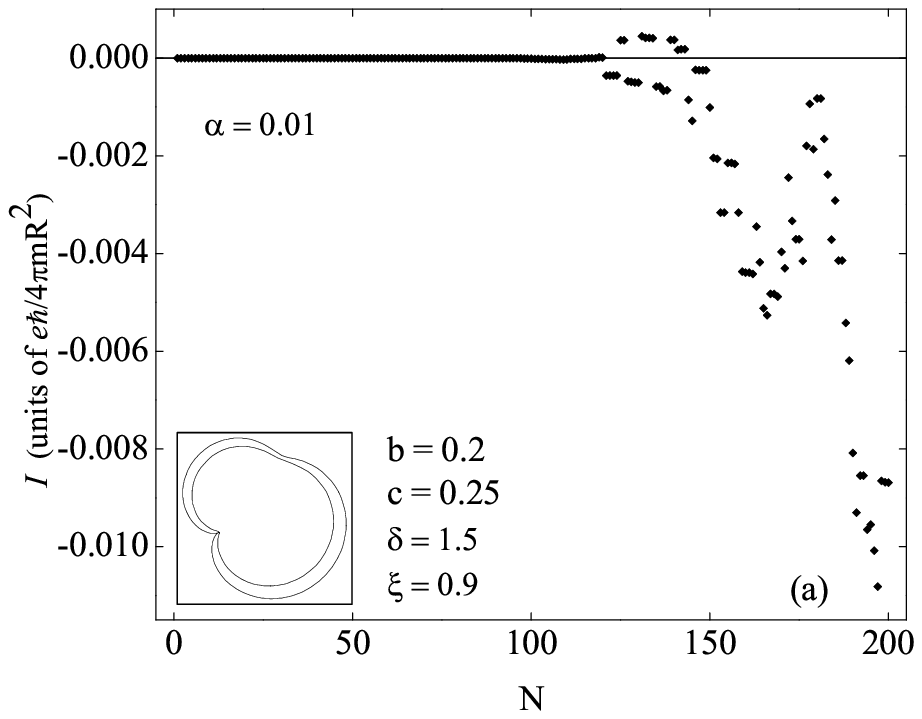}}
    \mbox{\includegraphics[width=3.30in]{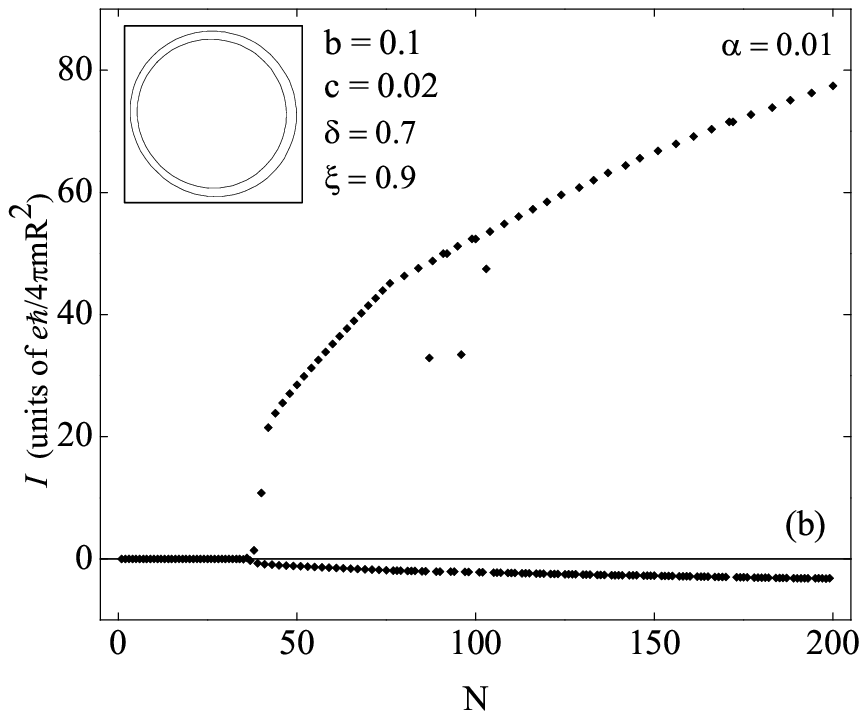}}
  }
  \caption{Persistent current $I$ vs. number of particles $N$ for distorted annulus. Parameters 
           b, c, and $\delta$ control the shape of billiard.}
  \label{I_vs_N_xi_0.9_chaot_ANNULUS}
  \end{figure*}

%-----------------------------------------------------------------------------

  For larger $\xi$ the magnitude of diamagnetic contribution to the
persistent current decreases, and the numerics are dominated by
mesoscopic fluctuations. When $\xi\to1$, the persistent current
becomes negligible (see Fig. \ref{I_vs_N_xi_0.9_chaot_ANNULUS}a) for
low occupations. We believe this is a manifestation of Anderson
localization due to the boundary scattering. At high energies, when
the localization length exceeds the circumference of the annulus,
extended states reappear and can carry persistent current. In
Fig. \ref{I_vs_N_xi_0.9_chaot_ANNULUS} we plot the persistent current
for 2 different sets of parameters controlling the shape of
annulus. For a large distortion
(Fig. \ref{I_vs_N_xi_0.9_chaot_ANNULUS}a) the current is nonzero only
for high energy states beyond $N=110$. In
Fig. \ref{I_vs_N_xi_0.9_chaot_ANNULUS}b the parameters $b, c$, and
$\delta$ are chosen to make the annulus less distorted, and we see
that the threshold for extended states moves to lower energy (about
$N=40$).

%%%%%%%%%%%%%%%%%%%%%%%%%%%%%%%%%%%%%%%%%%%%%%%%%%%%%

\section{Conclusions, caveats, and open questions}\label{conclusion}

%%%%%%%%%%%%%%%%%%%%%%%%%%%%%%%%%%%%%%%%%%%%%%%%%%%%%

We have investigated the behavior of  chaotic simply connected and
annular billiards penetrated by a pointlike flux. The annular
billiards are characterized by a dimensionless aspect ratio $\xi=r/R$, the
ratio of the inner ($r$) to the outer radius ($R$). Note that in the annular
billiards, the flux exists in a region where the electrons cannot
penetrate, and the effects of the flux on the electrons are purely
Aharanov-Bohm quantum intereference effects. Our main result is that
there is a systematic diamagnetic contribution to the persistent
current which can be traced back to the flux response of the $l=0$
levels of a regular unit disk (or annulus). Even though the number of
such $l=0$ levels is submacroscopic ($\propto\sqrt{N}$, where $N$ is
the number of electrons), the contribution to the persistent current
due to these levels is proportional to $N$ and is independent of the
flux for small flux in simply connected billiards, and can overwhelm
the fluctuating mesoscopic
contribution\cite{Cheung_1988,Cheung_1989,Ambegaokar_1990,Ambegaokar2_1990,Eckern_1991,Ambegaokar_1991,Schmid_1991,Oppen_1991,Altshuler_1991,Oppen_1993,Jalabert_1996,Richter_1996,Kawabata_1999}
from the states in the Thouless shell ($|E-E_F|\le E_T$). This effect
is quite distinct from Landau diamagnetism\cite{Nakamura_1988}.

The diamagnetic contribution to the persistent current from $l=0$
states seems to have been missed in previous work using the
semiclassical sum over periodic
orbits\cite{Oppen_1993,Jalabert_1996,Richter_1996,Kawabata_1999}. This
is understandable, since the semiclassical approach becomes exact only
as $E\to\infty$, and in this limit, the $l=0$ states have
vanishing spectral density $\rho_{l=0}(E)\simeq 1/\sqrt{E}$. However,
we emphasize that the total persistent current contains the sum over
all levels, and will indeed behave diamagnetically at small flux (in
the simply connected billiard) as we have described.

For very tiny $\xi$, the annular Robnik-Berry billiard behaves much
like the simply connected one for most values of the dimensionless
flux $\alpha=\Phi/\Phi_0\gg 1/\log{N\xi^{-2}}$, with a diamagnetic
contribution to the persistent current which is proportional to the
electron density. However, convergence to the $\xi=0$ limit is
logarithmically slow, and the limits $\alpha\to0$ and $\xi\to0$ do not
commute. As the aspect ratio $\xi$ increases, and the annulus tends to
a one-dimensional ring, this effect diminishes to zero. For $\xi$
close to $1$, we also see Anderson localization in the distorted
annular billiards, wherein the persistent currents are negligible
below a certain energy (presumably because the localization length for
these levels is smaller than the circumference), and become nonzero
only beyond a threshold energy.

While we can obtain analytical estimates for the limits $\xi\to0$ and
$\xi\to1$, it is difficult to make analytical progress for generic
values of $\xi$ (not close to $0$ or $1$). However, one can easily
verify from the asymptotic expansions that for generic $\xi$ the
diamagnetic contribution to the persistent current for $\Phi\ll\Phi_0$
goes as
\begin{equation}
I_{dia}\simeq -{\hbar^2\over \pi mrR} \alpha \sqrt{{2N(R-r)\over(R+r)}}
\end{equation}
where $r,\ R$, are the inner and outer radii respectively. This should
be compared to the typical fluctuating persistent
current for interacting particles\cite{Ambegaokar_1990,Ambegaokar2_1990,Eckern_1991,Ambegaokar_1991}
which behaves as
\begin{equation}
I_{fluc}\simeq {E_T\over\Phi_0}\simeq {\hbar^2\over mR\Phi_0}\sqrt{{N\over R(R-r)}}
\end{equation}
It can be seen that the ratio of the systematic diamagnetic persistent
contribution to the fluctuating contribution is roughly 
\begin{equation}
{|I_{dia}|\over|I_{fluc}|}\simeq {(R-r)\over r}{\Phi\over\Phi_0}
\end{equation}
Previous ring
samples\cite{Altshuler_1982,Webb_1985,Chandrasekhar_1985,Timp_1987,Wees_1989,Chandrasekhar_1991,Deblock_2002,Deblock2_2002}
have $(R-r)\ll r$. They are also subject to a uniform magnetic field
rather than a point flux. Despite this, a systematic diamagnetic
contribution at low flux has been detected in recent
experiments\cite{Deblock_2002,Deblock2_2002}. However, the experiments
are carried out at finite frequency, and the effects of attractive
pair interactions\cite{Schechter_2003,Murthy3_2004} (see below) or
nonequilibrium noise\cite{Kravtsov_2000} cannot be ruled out.

In order to detect this effect unambiguously, one must work with a
material which has no superconductivity at any temperature, to rule
out attractive pair interactions. It is also clear that ${R-r\over
r}$ needs to be made as large as possible in order to render this
effect easily observable. Care must be taken that there is no magnetic
flux in the region where the electron wavefunctions are nonzero in
order to maintain the pure Aharanov-Bohm quantum interference nature
of this effect.

Let us now mention some caveats about our work. We have taken only a
few ($\approx 200$) levels into account, whereas most experimental
samples have a hugely greater number of levels. However, the physics
of the diamagnetic contribution to the persistent current for a
particular level concerns only whether that level has $l=0$ or not,
and is independent of its relative position in the spectrum. We expect
our conclusions to hold for arbitrary densities.

We have considered a pointlike flux, which is unachievable in
practice. For the annular billiard, all one needs to ensure is that
the flux is nonzero only in the central hole of the annulus, and is
zero in regions where the electron density is nonzero. By gauge
invariance, such a situation will be equivalent to the one we study.

We have also ignored the effect of interelectron interactions. For
weak {\it repulsive}
interactions\cite{Andreev_1998,Brouwer_1999,Baranger_2000,Kurland_2000},
we expect interactions to modify the effect only slightly, because it
comes primarily from occupied levels deep within the Fermi sea, which
are Pauli-blocked from responding to the interactions. However, for
strong repulsive
interactions\cite{Murthy_2002,Murthy_2003,Murthy2_2004}, significant
corrections to the persistent current\cite{Murthy_2004} from electrons
in the Thouless shell cannot be ruled out. If the interactions are
weak but {\it
attractive}\cite{Ambegaokar_1990,Ambegaokar2_1990,Eckern_1991,Ambegaokar_1991},
the low-energy fluctuations of Cooper pairs become very
important\cite{Schechter_2003,Murthy3_2004}, and can produce
additional large diamagnetic contributions at low fields.

Similarly, though we have concentrated on the zero-temperature
behavior, we expect this effect to persist to quite high temperatures,
since most of the $l=0$ levels involved lie deep within the Fermi sea.

Finally, it would be interesting to investigate the effects of static
disorder within a chaotic billiard, which would induce the system to
cross over from a ballistic/chaotic to a disordered (diffusive)
system. We hope to address this and other issues in future work.

%%%%%%%%%%%%%%%%%%%%%%%%%%

  \begin{acknowledgments}

%%%%%%%%%%%%%%%%%%%%%%%%%%

    The authors would like to thank National Science Foundation for
    partial support under DMR-0311761 and DMR-0703992.  We are also
    grateful to Denis Ullmo for helpful discussions, and to Shiro
    Kawabata for bringing a missed reference to our attention. OZ
    wishes to thank the College of Arts and Sciences and the
    Department of Physics at the University of Kentucky for partial
    support, and GM thanks the Aspen Center for Physics for its
    hospitality, where part of this work was carried out.

  \end{acknowledgments}

%%%%%%%%%%%%%%%%%%%%%%%%%%%%%%%%%%%%%%%%%%%%%%%%%%%%%%%%%%%%

%---------------- BIBLIOGRAPHY ----------------------------

%\bibliographystyle{apsrev}
%\bibliography{paper}% Produces the bibliography via BibTeX.

%----------------------------------------------------------

%%%%%%%%%%%%%%%%%%%%%%%%%%%%%%%%%%%%%%%%%%%%%%%%%%%%%%%

%--------------------- APPENDIX ----------------------- 

\appendix

\section{Numerics for energy levels}\label{apnx:A}

The idea of this method is as follows
\cite{Robnik_1984,paper_Berry_Robnik, paper_Bruus_Stone_prb,
paper_Bruus_Stone_arxiv}.  In the original $(uv)$ domain the
Schr\"{o}dinger equation is:
\begin{equation}
 \frac{1}{2m}\left (-i\hbar {\bf \nabla} -q{\bf A}(u,v) \right )^2 \Psi(u,v) = E\Psi(u,v),
    \hspace{0.7 cm} q = -e < 0.
\end{equation}

 To keep dynamics of electron unchanged, it is assumed that
magnetic field exists only at the origin of $(uv)$ plane inside the billiard.
This requires that vector potential satisfies the condition
${\bf \nabla} \times {\bf A}({\bf r}) = {\bf n} \Phi \delta({\bf r})$, where ${\bf n}$
is a unit vector perpendicular to the plane of the billiard.

The billiard is threaded by single magnetic flux tube. The strength of the flux is
$\Phi = \alpha\Phi_0$, where $\Phi_0 = h/e$ is a magnetic flux quantum.

If the vector potential has the form:
\begin{equation}
 {\bf A}(u,v) = \frac{\alpha}{2\pi} \Phi_0 \left ( \frac{\partial f}{\partial v},  
                                                  -\frac{\partial f}{\partial u}, 0
                                           \right ), \ \ 
                 f = \frac{1}{2} \ln|z|^2,
\end{equation}
  then with the help of conformal transformation:
\begin{equation}
  w(z) = \frac{z + bz^2 + ce^{i\delta} z^3}{\sqrt{1 + 2b^2 + 3c^2}}, \ \ w = u+iv, \ \ 
                                                                         z = x+iy.
\label{eq:01}
\end{equation}
  the Schr\"{o}dinger equation in polar coordinates of $(xy)$ plane becomes:
\begin{equation}
 {\bf \nabla}^2_{r,\theta} \Psi(r,\theta) - \frac{i2\alpha}{r^2} \partial_{\theta}\Psi(r,\theta)
       - \frac{\alpha^2}{r^2} \Psi(r,\theta) 
       + \epsilon \left | w^{'}(re^{i\theta}) \right |^2 \Psi(r,\theta) = 0.
\label{eq:02}
\end{equation}
 Here the energy $\epsilon$ is measured in units of $\hbar^2/2mR^2$, and the distance is in units
of $R$, where $R$ is the radius of the disk in $(xy)$ plane. Also, the coefficients $b,c$,
and $\delta$ in Eq. \eqref{eq:01} are real parameters selected in the way so that $|w^{'}(z)| > 0$
for all values of $z$ inside the disk in $(xy)$ plane. The transformation $w(z)$ is a cubic
polynomial normalized to preserve the area of the billiard and leave the density
of states invariant.  Equation \eqref{eq:02} should be accompanied by Dirichlet boundary
condition.

  To find the energy spectrum, one expands the $\Psi(r,\theta)$ function in Eq. \eqref{eq:02} in
terms of the eigenstates $\phi_{l,n}(r,\theta)$ of free electron($w=0 $) inside the round
billiard ($R=1$):
\begin{equation}
 \Psi_p(r,\theta) = N_p \sum^{\infty}_{j=1} \frac{c^{(p)}_j }{\gamma_j} \phi_j(r,\theta).
\label{eq:03}
\end{equation}

\noindent Compound index $j=(\nu,n)$ numerates levels in ascending order. Normalized function
 $\phi_{l,n}(r,\theta)$ is:
\begin{equation}
 \phi_{l,n}(r,\theta) = \frac{J_{|l-\alpha|}(\gamma_{|l-\alpha|,n}r) e^{il\theta}}
                             {\sqrt{\pi} J^{'}_{|l-\alpha|}(\gamma_{|l-\alpha|,n})}
\end{equation}
  Function $J_{\nu}(r)$ is the Bessel function of the first kind, $\gamma_{\nu,n}$ is the $n$th
root of $J_{\nu}(r)$, and $l$ is an orbital quantum number. The coefficients of expansion in
Eq. \eqref{eq:03} are chosen that way for further convenience.

  Plugging expansion \eqref{eq:03} into Eq. \eqref{eq:02}, after simplification one gets the matrix
equation for eigenvalue problem:
\begin{equation}
 M_{ij} c^{(p)}_j = \frac{1}{\epsilon_p} c^{(p)}_i,
\end{equation}
  where matrix $M$ is:
\begin{equation}
 \begin{split}
  M_{ij} = \biggl[ \frac{\delta_{ij}}{\gamma_i\gamma_j} 
                &+ \delta_{l_i,l_j-2} 6ce^{-i\delta} I^{(2)}_{ij}
                + \delta_{l_i,l_j-1} (4bI^{(1)}_{ij} + 12bce^{-i\delta} I^{(3)}_{ij} )\\
                &+ \delta_{l_i,l_j} (8b^2I^{(2)}_{ij} + 18c^2 I^{(4)}_{ij} )
                + \delta_{l_i,l_j+1} (4bI^{(1)}_{ij} + 12bce^{i\delta} I^{(3)}_{ij} ) \\
                &+ \delta_{l_i,l_j+2} 6ce^{i\delta} I^{(2)}_{ij} \biggr]/(1+2b^2+3c^2).
 \end{split}
\end{equation}
  The integrals $I^{(h)}_{ij}$ have the form:
\begin{equation}
 I^{(h)}_{ij} = \frac{\int_0^1 dr r^{h+1} J_{\nu_i}(\gamma_ir) J_{\nu_j}(\gamma_jr)}
                     {\gamma_i \gamma_j J^{'}_{\nu_i}(\gamma_i) J^{'}_{\nu_j}(\gamma_j)}.
\end{equation}

Along with the simply connected domain (irregular disk) we consider irregular annulus.
Using similar conformal transformation, we map the annulus with irregular boundaries
from $(uv)$ plane onto regular annulus in $(xy)$
plane with inner radius $\xi$ and outer radius $R=1$. Proper conformal transformation looks as
follows:
\begin{equation}
  w(z) = \frac{z + bz^2 + ce^{i\delta} z^3}
              {\sqrt{1 + 2b^2(1+\xi^2) + 3c^2(1+\xi^2+\xi^4)}}.
\end{equation}

  For this kind of billiard expansion of $\Psi(r,\theta)$ from Eq. \eqref{eq:03} is in terms of
eigenstates $\phi_j(r,\theta)$ for regular annulus:
\begin{equation}
 \phi_{l,n}(r,\theta) = \frac{\left[ J_{|l-\alpha|}(\gamma_{|l-\alpha|,n}r) 
                        - \dfrac{J_{|l-\alpha|}(\gamma_{|l-\alpha|,n}\xi)}
                               {N_{|l-\alpha|}(\gamma_{|l-\alpha|,n}\xi)}
                        N_{|l-\alpha|}(\gamma_{|l-\alpha|,n}r) \right] e^{il\theta}}
                             {\sqrt{2\pi} \sqrt{\mathlarger{\int}_{\xi}^1 dr r \left[  
                              J_{|l-\alpha|}(\gamma_{|l-\alpha|,n}r)
                                  - \dfrac{J_{|l-\alpha|}(\gamma_{|l-\alpha|,n}\xi)}
                                       {N_{|l-\alpha|}(\gamma_{|l-\alpha|,n}\xi)}
                             N_{|l-\alpha|}(\gamma_{|l-\alpha|,n}r) \right]^2 }}
\label{eq:annul_wave_func}
\end{equation}

  The counterpart of matrix $M$ for the annulus is:
\begin{equation}
 \begin{split}
  M_{ij} = \biggl[ \frac{\delta_{ij}}{\gamma_i\gamma_j}
                &+ \delta_{l_i,l_j-2} 3ce^{-i\delta} I^{(2)}_{ij}
                + \delta_{l_i,l_j-1} (2bI^{(1)}_{ij} + 6bce^{-i\delta} I^{(3)}_{ij} )\\
                &+ \delta_{l_i,l_j} (4b^2I^{(2)}_{ij} + 9c^2 I^{(4)}_{ij} )
                + \delta_{l_i,l_j+1} (2bI^{(1)}_{ij} + 6bce^{i\delta} I^{(3)}_{ij} ) \\
                &+ \delta_{l_i,l_j+2} 3ce^{i\delta} I^{(2)}_{ij} \biggr]/(1 
                      + 2b^2(1 + \xi^2) + 3c^2(1 + \xi^2 + \xi^4)).
 \end{split}
\end{equation}
  where the integrals $I^{(h)}_{ij}$ are defined as:
\begin{equation}
  I^{(h)}_{ij} = \int_{\xi}^{1} dr r^{h+1} \frac{\tilde{\phi}_i(r) \tilde{\phi}_j(r)}
                                                {\gamma_i \gamma_j}, \ \ \ 
                 \tilde{\phi}_i(r) = \sqrt{2\pi} e^{-il\theta} \phi_i(r,\theta).
\end{equation}

%%%%%%%%%%%%%%%%%%%%%%%%%%%%%%%%%%%%%%%%%%%%%%%%%%%%%%%%%%%%

%---------------- BIBLIOGRAPHY ----------------------------

%%%%%%%%%%%%%%%%%%%%%%%%%%%%%%%%%%%%%%%%%%%%%%%%%%%%%%%%%%%%

\bibliographystyle{apsrev}
\bibliography{paper}% Produces the bibliography via BibTeX.

\end{document}